\documentclass{IEEEtran}
\usepackage{amsmath,amsthm}
\usepackage{authblk}
\usepackage{algorithmic}
\usepackage{algorithm}
\usepackage{stmaryrd}
\usepackage{pxfonts}
\usepackage{graphicx}
\usepackage{array}
\title{A Reputation Based Framework to Avoid Free-riding in Unstructured Peer-to-Peer network}
\author{Ruchir Gupta, Yatindra Nath Singh, \emph{Senior Member IEEE}}
\affil{Department of Electrical Engineering, IIT, Kanpur}
\affil{\textit {\{rgupta,ynsingh\}@iitk.ac.in}}
\begin{document}

\maketitle
\begin {abstract}
Free riding is a major problem in peer-to-peer networks. Reputation management systems are generally proposed to overcome this problem. In this paper we have discussed a possible way of resource allocation on the basis of reputation management system i.e. probabilistic allocation based on reputation. This seems to be a better way for allocation of resources because in this case nodes that don't have very good reputation about each other, may also serve each other at least some amount of resource with finite probability. This avoids disconnect between them. Algorithms are presented for optimizing the shared capacity, reputation based probabilistic allocation that is optimal for a node, and formation of interest groups on the basis of similarity between interests of nodes.
\end {abstract}

%\subsection{test}
%\chapter{Probabilistic Resource Allocation in Peer-to-Peer Networks}
\section {Introduction}
The reputation computation should be such that it gives enough credit to the nodes those are contributing more to the network. If some node is requested for smaller size resource, it should not be at loss. We are proposing a probabilistic resource allocation method in this paper. It uses reputation of users while allocating the resources. This kind of allocation ensures that a node with low reputation gets resources with some small finite probability. This will avoid the disconnect between two nodes that have low reputation for each other. This is important when resources are distributed in the network.

The incentive systems proposed in literature do not consider the interests of peers i.e. the neighbourhood of a peer is not made up on the basis of its interests. Although, gnutella servant keeps few last query replying peers in the cache but it is limited to that only. This system is rather inefficient because two nodes may have many interests in common while not being neighbours. Whereas two neighbour nodes may have very few common interests. Hence, the better option is to make neighbourhood on the basis of similarity of interests and reputation. 

In this paper, we have proposed an algorithm for the optimization of shared capacity of a node (section \ref{CS5}), a method to compute the reputation (section \ref{RBS5}), probabilistic resource allocation based on reputation (section \ref{RBS5}) and server selection according to interests of node and reputation (section \ref{SS5}).
Finally numerical results are given to verify the hypotheses (section \ref{NR5}).
\section{Related Work}
\label{RW5}
Various groups have suggested different techniques for resource allocation in peer-to-peer networks. Kung \emph{et.al.} \cite{Kung03differentiatedadmission} proposed selection of a peer for allocation of resource according to its contribution to the network and usage of resources. In the same context, Feldman \emph{et.al.} \cite{Feldman2004} proposed a new term -- generosity of the node. It is estimated as the ratio of the service provided by the node to the service received by the node. Nodes will be served as per their estimated generosity. Banerjee \emph{et.al.} in \cite{reciprocal} proposed that a node will calculate the expected utility function for requesting node and on that basis it will decide if service has to be provided or not. In \cite{Satsiou:2010,rationalmodel} the resource allocation algorithm for single link limited capacity systems has been proposed. These papers considers network as market and proposes that second price auction leads to optimality. \cite{Satsiou:2010} assumes that resource is available everywhere except at the requesting node. Thus, a node is not required to have interaction with many nodes. Ma \emph{et.al.} \cite{Ma:2006} proposed progressive water filling algorithm on the basis of marginal utility for allocation of resources among different requesting nodes. The base of bucket for water filling is proposed to be varying according to the contribution of requesting node.  Ma \emph{et.al.} \cite{ma} proposed to allocate the resource to requesting node on the basis of their contribution and requirement of bandwidth. Yan \emph{et.al.} in \cite{ranking-basedoptimal} proposed a ranking based resource allocation scheme. Resource allocation is done according to utility and ranking of requesting peer to ensure max-min fairness.  

Social networks are formed on the basis of interests of users. This fact is been capitalised to improve query search as well as recommendation network in peer-to-peer networks \cite{social1, social2, Trust-aware-query, socp2p}. In \cite{social1} BitTorrent traces are studied and it is concluded that interest based grouping of peers results in an efficient system. It also proposes a DHT based system to implement this kind of group formation. Wang \emph{et.al.} \cite{Trust-aware-query} proposed interest based online social communities that are headed by super nodes and nodes join the communities according to their interests. These communities will have a trust relationship among its members. In \cite{socp2p} a friend network is proposed on the basis of similarity of interests.
%\section{System Model}
\section{Capacity Sharing}
\label{CS5}
As every node in the network is rational, hence it will try to share minimum amount of resource to increase its pay-off. If there is a reputation management system implemented in the network, nodes are compelled to share the resources to get the quality of service from the network. Nodes, being rational, will try to optimize the amount of shared resources. We propose a method for nodes to optimize the shared capacity to get the required quality of service from network. In this method, nodes will initially share some amount of resource. This amount of shared resource will be periodically reviewed and adjusted for optimality.

Initially a node will share the capacity as per its perceived download requirement. By perceived download requirement, we mean a rough estimate of its average download requirement. This need not to be accurate as it will be updated later on. But, it should neither be too low to ruin the reputation of a newcomer node nor be too high to cause a cost penalty. If no estimate is available, initially half of total download capacity can be shared.

Node will tweak the value of its shared capacity by periodically increasing and decreasing it by some amount $\delta$ to get the optimal point where it will get maximum advantage. While doing so, node will follow the following method.
\begin{enumerate}
\item If decrease in sharing capacity does not decrease significant average download, it implies that node is sharing more than required resource and hence it should decrease it.
\item If increase in sharing capacity increases significant average download, it implies that node is sharing less than required resource and hence it should increase it.
\item If decrease in sharing capacity decreases significant average download, it implies that either node was on optimal point (if it was preceded by an increase) so it should get back to that point or is now sharing even lesser than what it should have shared.
\item If increase in sharing capacity does not increases significant average download, it implies that either node was on optimal point (if it was preceded by an decrease) or is now sharing even more than what is required.
\end{enumerate}

\begin{algorithm}
\caption{Shared upload capacity adjustment of a node}
\label{scalgo}
\begin{algorithmic}
\STATE   $k=0$; and $A(k)=-1$
\COMMENT {k is the instant when node reviews its sharing capacity and A(k) is the indicator variable which shows the action taken at a particular k}
\REPEAT
\STATE   $D_k$ $\shortleftarrow$ average data download for $kT$ to $(k+1)T$
\STATE $U_s=U_s+\delta\cdot A(k)$
\COMMENT {$U_s$ is shared capacity of the node}
%\STATE   $D_{k+1}$ $\shortleftarrow$ average data download for $(k+1)T$ to $(k+2)T$
\IF {$|D_k - D_{k-1}| \le \epsilon$}  
\IF {$A(k)= A(k-1)$ or A(k)=0}
\STATE $A(k+1)\shortleftarrow -1$
\ELSE
\IF {$A(k)=1$ \&\&  $A(k-1)=-1$}
\STATE \STATE $A(k+1)\shortleftarrow 0$
\ENDIF
\ENDIF
\ELSE
\IF {$A(k)=-1$ \&\&  $A(k-1)=1$ \&\& $D_k<D_{k-1}$}
\STATE \STATE $A(k+1)\shortleftarrow 1$
\ENDIF
\IF {$D_k> D_{k-1}$}
\STATE $A(k+1)\shortleftarrow 1$
%\ELSE 
%\STATE $A(k+1)\shortleftarrow 0$
\ENDIF
\ENDIF
\STATE $k\shortleftarrow \mod_5(k+1)$  
\UNTIL Node is in the network
\end{algorithmic}
\end{algorithm}
 This process is shown in algorithm \ref{scalgo}.

$\epsilon$ is a parameter that is kept for overcoming the effect of demand variation in the network. When node observes high variation of the demand in the network, value of $\epsilon$ will be increased.
\section{Reputation Based System}
\label{RBS5}
Network is only meaningful if nodes are interacting with each other and contributing to each others' interest. We have a network of nodes that are rational in nature. Such nodes contribute in the network only when they have some incentive for doing so. To avoid this problem, a reputation based incentive system can be used. In such a system nodes keep the record of behaviour o
f other nodes observed by itself or on the basis of recommendation of different nodes. This kind of system forces rational nodes to contribute to the network. To implement this kind of a system, we need to formulate a way for estimation of reputation and a way for allocation of resource according to the estimated reputation.
\subsection{Reputation Management System}
Ideally, reputation should be the measure of cooperative behaviour of a node which is an abstract quantity and it is a private information of a node. So, it is difficult to measure the cooperative behaviour of a node and we can only measure its implications with some degree of uncertainty. However, it can be estimated with certain accuracy on the basis of behaviour observed by a node.

There could be a number of ways to observe the behaviour of a node. One such method may be to use the ratio of received data rate to requested data rate. The advantage of such technique is that if some node is asking for less amount of data, the serving node will not earn a bad reputation. Moreover this kind of reputation remains between $0$ and $1$ as given in \ref{SM2},
 \cite{ruchirestm}, i.e.,
\begin{equation}
\label{repeq}
t_{ij}=\left(\frac{q_{a,ji}}{min(q_{ij,ay},q_{f,ji})}\right)^{1-\eta_i} \times \frac{\hat{q}_{w,ji}}{q_{r,ji}},
\end{equation}
hence it is easy to handle. Here $t_{ij}$ is the reputation of node $j$ for node $i$, $q_{r,ij}$,  $q_{w,ji}$, $q_{a,ji}$, $q_{f,ji}$ and $q_{ay,ji}$ are the requested, willing, actual, feasible and accepted service rates respectively.  
The disadvantage of this kind of system is that it does not takes into account the amount of request. It means that if a node is asking for less amount of resource from a node and more amount of resource from another node and both are fulfilling node's demand, both will get similar gain in reputation. However, the node that was requested more resource, had to pay more in comparison to the other one. 

This problem can be taken care of by giving different weights to different transactions as per the amount of resource requested by that node. Weights should be such that these should range between 0 to 1  and  biggest service request should get maximum weight. Requesting node has a fixed download capacity that is generally the maximum of its download requirement.

A node can calculate the reputation of a node with following formulation,
\begin{equation}
\label{9}
t_{ij}=\left(\frac{q_{a,ji}}{min(q_{ij,ay},q_{f,ji})}\right)^{1-\eta_i} \times \frac{\hat{q}_{w,ji}}{q_{r,ji}}\times \frac{q_{r,ji}}{q_{r,i,d}}.
\end{equation}
Here $q_{r,i, d}$ is the download capacity of node $i$. It may be noted that we have multiplied the factor $\frac{q_{r,ji}}{q_{r,i,d}}$ to the $t_{ij}$ as estimated by equation \ref{8} in paper \ref{chap2}.

Keeping the download capacity ($q_{r,i,d}$) instead of requested resource ($q_{r,ji}$) in denominator, we can overcome the above mentioned problem. However, $q_{r,i,d}$ is quite large compared to $q_{r,ji}$, this will make reputation values very low. Apart from it, every node has a different value for $q_{r,i,d}$ and it will be a problem for aggregation because a node having same kind of behaviour with two nodes of different download capacities will have different value for the $t_{ij}$. To over come these problems nodes will multiply their reputation table with $({q_{r,i,d}}/{Q_{r,d}})$. Here $Q_{r,d}$ is the universal scaling factor known by all the nodes.

\label{scalingprob}
However, this will only work well if a node is only interested in resources that have size of same order. If some node 'A' is rich with resources that are small in size, this node will have a very small value of reputation for a node that is requesting for all size of resources because 'A' would have been asked only for small amount of resource. Consequently, when 'A' asks for small amount of resource, it will get the resource with a very small probability. This issue will be further investigated in next subsection.
\subsection{Probabilistic Resource Allocation}
\label{allocation}
In probabilistic resource allocation, node probabilistically decides if it will provide the resource to requesting node or not. It means, when a node 'A' requests for resource from node 'B', node 'B' checks the reputation table and converts reputation of 'A' to its effective reputation. Here, by effective reputation we mean reputation that is adjusted according to the requested amount of resource. To calculate this value, node multiplies the reputation value with ratio of its download capacity to requested amount of resource, i.e.
\begin{equation}
t_{ij, effective}= t_{ij}\times \frac{q_{r,i,d}}{q_{r,ij}}.
\end{equation}
 Now in proposed system, if a node is asking similar amount of resource as it supplied, it will be given same quality of service. If it asks for smaller resource than it supplied, it gets even better quality of service whereas if opposite happens, it gets a poor quality of resource.

Once node $i$ gets the effective reputation of a node $j$, it selects the node $j$ with probability proportional to its reputation. It means node $i$ generates a random number. If this generated number is smaller than the reputation of requesting node i.e. node $j$ here, multiplied by a constant ($\nu_i$), requesting node is selected to provide the resource. $\nu_i$ is the constant that ensures the requirement of selected nodes remains around the shared capacity so that it can be optimally utilised. Mathematically,
\begin{equation}
P_{allocation,i,j}=\begin{cases}\acute{P}_{allo} &\text{if $\acute{P}_{allo}< 1$}. \\
	 1, &\text{otherwise}.
	\end{cases}
\end{equation}
Where,
\begin{equation}
\nonumber\acute{P}_{allo}=(t_{ij, effective})^x\cdot \nu_i
\end{equation}
Here $P_{allocation,i,j}$ is the probability by which $j$ will be allocated the resource. Node will be selected for resource allocation if,
\begin{equation}
\label{nodeselection}
rand\le P_{allocation,i,j}
\end{equation}
Here $rand$ is the random number generated by the node and x is the reputation exponent. It is used because a low reputation node can only increase its reputation if it serves with $x<1$. Its value has been calculated in \cite{ruchirwhite}.

Nodes will dynamically and periodically adjust the value of  $\nu_i$ to get the optimality. To do so, node will measure the utilised part of its shared capacity and fulfilment level of demand of selected nodes. If it is not able to utilise its shared capacity regularly, it increases the value of $\nu_i$ and if demand of selected node is not fulfilled over the time, value of $\nu_i$ is decreased.

After the selection of nodes, the shared capacity is distributed among selected nodes. If the total demand of requesting nodes is less than the shared capacity of serving node, every node is allocated resource as per their requirement. If this total demand is greater than the shared capacity, node needs to use some kind of allocation algorithm such that, the serving node can get maximum advantage and nodes can not play game by asking for resources greater than their requirement.

A node will be maximum benefited when it has highest chance of getting selected for resource allocation i.e. by maximizing its $P_{allocation}$ for a future time when it will need some resource, according to equation (\ref{9}). If serving node is doing this calculation, it can be assumed that $q_{a,ij}=min(q_{ji,ay},q_{f,ij})$. Hence equation reduces to,
\begin{equation}
\label{5}
t_{ji,effective}= \frac{\hat{q}_{w,ij}}{q_{r,ij}}\times \frac{q_{r,ij}}{q_{r,ji}}.
\end{equation}
$\hat{q}_{w,ij}$ can be replaced by $q_{o,ij}$ as we are discussing about allocation in a particular round where number of nodes, their demands and shared capacity has already been fixed. Hence equation (\ref{5}) reduces to,
\begin{equation}
 t_{ji,effective}= \frac{q_{o,ij}}{ q_{r,ji}}.
\end{equation}
$q_{r,ji}$ can only be predicted statistically. We can observe that if a node 'A' is asking less amount of resource to node 'B' then 'B' can only get less amount of resource because if it will demand for bigger resource, its effective reputation will come down and hence it will not be selected for service by node 'A'. Hence if A is asking lesser resource from B, that implies B is asking lesser resource from A.

For simplicity $\nu$ can be taken as 1. Therefore, the optimisation problem a node needs to solve becomes,
\begin{equation}
\max\sum\limits_j (t_{ji,effective})^x\implies \max\sum\limits_j \left(\frac{q_{o,ij}}{ q_{r,ij}}\right)^x.
\end{equation}
Such that
\begin{eqnarray}
\nonumber\sum\limits_j q_{o,ij}&=& U_{s,i}\\
\nonumber q_{o,ij}&\le & q_{r,ij}; \;\;\; \forall j\\
\nonumber\sum\limits_j q_{r,ji}&>& U_{s,i}
%\nonumber x &=& constant \in (0,1)
\end{eqnarray}
Here $x$ is a constant. Its value lies between 0 and 1. For our case its value is 0.75.

 This is a difficult optimisation problem. For that, we need to observe the function $a^x$. Here $a$ varies from 0 to some finite value and $x$ is a constant between 0 and 1 as mentioned above. Two facts are easy to observe, first it is a monotonically increasing function and second its rate of change is monotonically decreasing function. Therefore it is evident that initially our objective function will get the maximum increment if resource is allocated to node that corresponds to smallest $q_{r,ij}$. After some allocation, increase in the value of objective function will decrease and now it will be more for any other node that has requested more data than first one. Now it will be beneficial to allocate data to this second node. After some allocation, any other node may result in more increment and this process continues till the resource allocation is complete.

On the basis of reason mentioned above, we propose an algorithm for allocation. First a node $i$ decides about the minimum unit of allocation. Let us call this $\Delta_i$. On the basis of $\Delta_i$ and the amount of total resource shared ($U_{s,i}$), $i$ calculates the total number of allocation units ($U_{su,i}$) such that $U_{su,i}=\frac{U_{s,i}}{\Delta_i}$. Now, $i$ constructs an allocation array $U_{sua}$ that has the dimension of $U_{su,i}\times \acute{N}$. Here $\acute{N}$ is the number of nodes selected for data allocation by equation \eqref{nodeselection} such that
\begin{equation}
U_{sua}(k,j)=((k)^x-(k-1)^x)\cdot\left(\frac{\Delta_i}{q_{r,ij}}\right)^x.
\end{equation}
Here $k$ and $j$ are row and column indices of $U_{sua}$. Elements of $U_{sua}$ are sorted and indices of top $U_{su,i}$ element are stored in a vector of dimension $1\times U_{su,i}$. The number of times any particular node comes in this vector will be allocated the same number of units.  

In this kind of allocation, nodes asking for less amount of data will be given data first. Hence if a node asks for more data than its requirement, it loses the allocation part. This kind of allocation will also fulfil our second requirement.

As requests will be coming temporally in arbitrary fashion, it is necessary to define a policy followed by a node for provisioning the service. If node will serve the request as and when it comes, node will always remain busy in doing so. Moreover, nodes that has got the bandwidth, will get allocation again and again. If node will service the requests periodically, there is a chance with finite probability that a low reputation node will get the service while a high reputation node may keep on waiting.

Hence, a node should have a dynamic policy about serving instants. It means that when total reputation of requesting nodes crosses a certain threshold, node will serve the accumulated requests. If over a certain period of time total reputation of requesting nodes does not cross the threshold, node will serve the requests accumulated by this time. While summing up the reputation of requesting nodes, it is ensured that high reputation nodes get preferred to the nodes of lower reputation. Whenever, a node serves new requests, node will first do the selection process for newcomer nodes and then it will redistribute the resources among newly selected nodes and already existing nodes.

\section{Server Selection}
\label{SS5}
\subsection{Common Interest Groups}
In peer-to-peer file sharing network, different users have common interests. For a user, it is beneficial to make neighbours that share interests with him and ready to serve him. Therefore, a node should adopt a strategy to form its neighbourhood according to similar interests with good reputation nodes.

Interest is an abstract notion so classification of nodes on the basis of interest is difficult. Even if it is done, this will be a very large set that will be difficult to handle. Therefore, interest group should be formed on the basis of files, users requested or provided. However, users with different interests may request same file. For example, a song may be liked for different reasons like music, singer or lyrics. But, if two users are requesting for more and more similar files, probably they may have some common interests. As the number of similar files grows, probability of two peers choosing file due to same interest increases while choosing it for different interests decreases.

Therefore, we propose that a node will compute the similarity coefficient of the other nodes in the network. The similarity coefficient ($\chi_{ij}$) of node $j$ will be calculated by node $i$ using
\begin{equation}
\chi_{ij}=\begin{cases}v_{ij} \cdot log_{base_i} (\Omega_{ij}+1) &\text{if $\Omega_{ij}< base_i$}. \\
	 v_{ij}, &\text{otherwise}.
	\end{cases}
\end{equation}
Here $\Omega_{ij}$ is the number of times node $i$ has queried to node $j$ or vice-versa, $v_{ij}$ is the ratio of answered queries to total queries between node $i$ and node $j$. $base_i$ will be dynamically adjusted periodically as per the accuracy of similarity coefficient of the node. It means if the selected neighbours can not answer sufficient number of requests, value of $base_i$ will be increased.
\subsection{Inclusion of Reputation in Neighbourhood Formation}
As discussed earlier, for server selection, a node need to form its neighbourhood using interests and reputation. This can be done by combining reputation ($t_{ij}$) and similarity coefficient ($\chi_{ij}$) for node $j$. The combined score can be used to rank the other nodes in the network. This rank can be used to select the server i.e. where to send queries.

The combining can be done as follows.
\begin{equation}
score_{ij}=\alpha\cdot \chi_{ij}+(1-\alpha)\cdot t_{ij}.
\end{equation}
Here $\alpha$ is a combination coefficient between 0 and 1. Value of $\alpha$ will depend upon the stability of common interest network. If a node has newly joined the network, it has to build the interest network hence $\alpha$ will be taken high. Once it has a stable interest network, value of $\alpha$ will be decreased to have more contribution of reputation in the score.

\section{Numerical Results}
\label{NR5}
\begin{figure}[!t]
\begin{center}
\includegraphics[width=85mm,height=85mm, keepaspectratio=false]{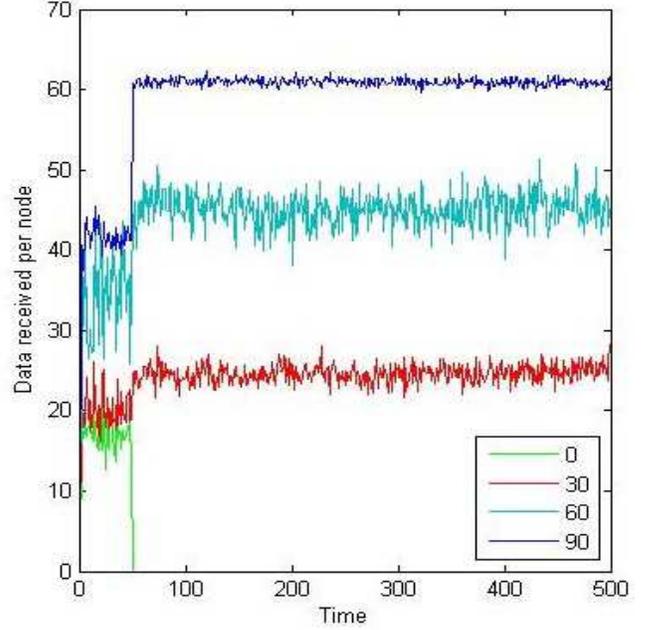}
\caption{Average data received per node within the same network. Different graphs show nodes with different shared capacities}
\label{datarecdpershare}
\end{center}
\end{figure}

\begin{figure}[!t]
\begin{center}
\includegraphics[width=85mm,height=85mm, keepaspectratio=false]{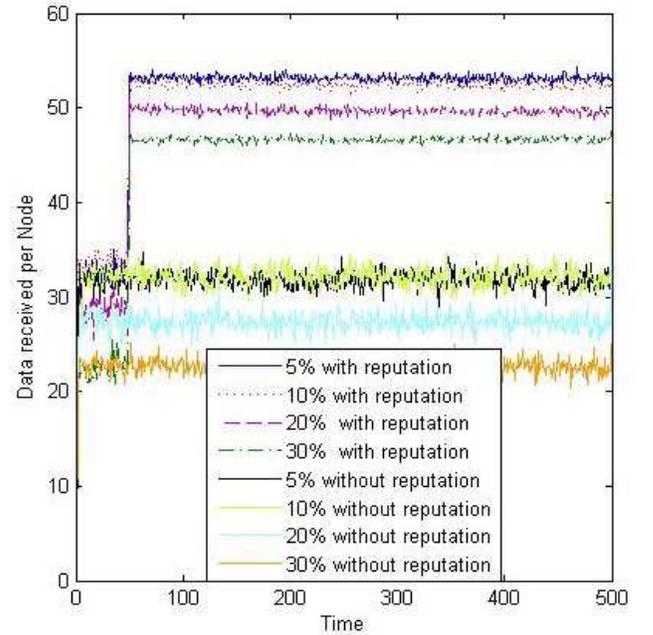}
\caption{System performance with different percentage of free riders.}
\label{freerider}
\end{center}
\end{figure}

\begin{figure}[!t]
\begin{center}
\includegraphics[width=85mm,height=85mm, keepaspectratio=false]{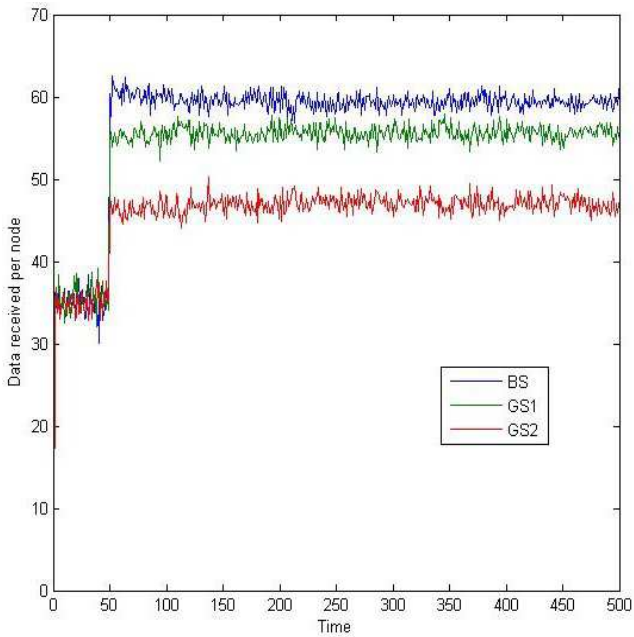}
\caption{Data received by nodes with different strategies}
\label{bsgs}
\end{center}
\end{figure}

\begin{figure}[!t]
\begin{center}
\includegraphics[width=85mm,height=85mm, keepaspectratio=false]{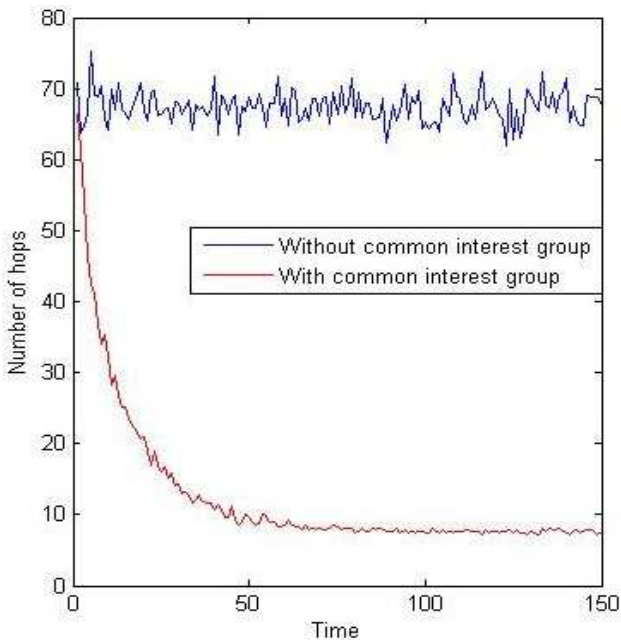}
\caption{Average number of nodes queried required for query resolution}
\label{interestgroup}
\end{center}
\end{figure}
We have done performance evaluation of reputation system and resource allocation system for a network of 200 nodes. We have also evaluated interest based group formation algorithm for a network of 1000 nodes. We have considered the discrete time instants for the purpose of measurement and estimation in the simulations. Every slot is termed as an iteration. First $50$ iterations have been taken as an acquaintance period i.e. a node will allocate their bandwidth without referring to the reputation table.

Figure \ref{datarecdpershare} presents the average data received by nodes sharing different amount of resource to the network. Here, it is evident from figure that the node that is sharing more data, is getting better quality of service. Figure \ref{freerider} shows the performance of system in presence of different percentage of free riders. We can see in figure that from 5\% to 10\% decay in system performance is almost negligible. After that, system performance decreases by small amount. So we can say that system performance does not deteriorate much due to free riders.

Figure \ref{bsgs} shows the data received by peers asking for different amount of data in the network. Here, BS represents the nodes that request for the amount of resource as per its requirement whereas GS1 and GS2 represents the nodes that requests the amount of resource multiple time to their requirement. GS2 requests more times than GS1. Here it can be seen that nodes making request as per their requirement are getting better quality of service whereas nodes that are trying to exploit network by making requests multiple times are not getting that kind of quality of service. This discourages the tendency of exploitation of by making multiple time requests.

Figure \ref{interestgroup} shows the average number of nodes queried required for resolution of query in interest based and non-interest based network. Here, it can be seen that, if node forms interest groups, its query gets resolved in much lesser number of hops than number of hops in other case.
\section{Conclusion}
\label{CON5}
In this paper, we have discussed allocation of resource by node on the basis of reputation. Allocation has been done probabilistically, i.e., requesting node has been offered resource with a probability proportional to its reputation. If total demand of selected nodes is more than offering node's shared capacity, allocation will be done to optimise the gain in reputation of offering node. An algorithm has been proposed for the same. This algorithm also ensures that nodes do not request more than their actual demand. An  algorithm for formation of common interest group and shared capacity optimisation has also been proposed. Numerical results show that proposed algorithms work as per the requirement.

\bibliography{ref}{}
\bibliographystyle{ieeetr}
\end{document}